\documentclass[a4paper]{iopart}
\usepackage{latexsym}
\usepackage[dvips]{graphicx}   
\usepackage{amsmath}           
\usepackage[all]{xy}           
\usepackage{amsfonts}          
\usepackage{amssymb}           

  \usepackage[english]{babel}

\newcommand{\ef}[1]{\, #1}     

\newcommand{\references}{
\section*{References}}

\newcommand{\modulo}[1]{\quad (\textrm{mod}\  #1 )}
\newcommand{\deltakr}{\delta
  }                                            

\newcommand{\reval}[1]{\overline{#1}}
\newcommand{\sspan}{\mathrm{span}} 
\newcommand{\kker}{\mathrm{ker}}
\newcommand{\rrank}{\mathrm{rank}}

\newcommand{\vett}[1]{{\bf #1}} 

\newcommand{\true}{\textrm{TRUE}}
\newcommand{\false}{\textrm{FALSE}}
\newcommand{\band}{\wedge}
\newcommand{\bor}{\vee}
\newcommand{\bxor}{\dot{\vee}}

\newcommand{\grid}{J}
\newcommand{\punto}{p}
\newcommand{\ham}{\mathcal{H}}
\newcommand{\Perm}{q\mathit{-Perm}}

\newcommand{\erre}[1]{\mathcal{R}_{#1}}
\newcommand{\gausspol}[2]{\genfrac{[}{]}{0pt}{}{#1}{#2}}

\newtheorem{ansatz}{Ansatz}[section]
\newtheorem{theor}{Theorem}

\newcommand{\myquote}[1]
{\begin{quotation}
\noindent
\it{``#1''}
\end{quotation}}

\begin{document}

\jl{1}

\def\PACS{\par\leavevmode\hbox {\it PACS: 89.80.+h, 75.10.Nr}}

\article {}{An exactly solvable random satisfiability problem}

\author
[
  Sergio Caracciolo,
  Andrea Sportiello
]
{
  Sergio Caracciolo$^{[1]}$,
  Andrea Sportiello$^{[2]}$
}
\address{[1] Universit\`a degli Studi di Milano - Dip. di Fisica and INFN,
\\ via Celoria 16, I-20133 Milano,  and NEST-INFM, Italy
\\ Mail address: \tt{Sergio.Caracciolo@mi.infn.it}}
\address{[2] Scuola Normale Superiore and INFN - Sez. di Pisa,
\\ Piazza dei Cavalieri 7, I-56100 Pisa, Italy
\\ Mail address: \tt{Andrea.Sportiello@sns.it}}

\date{\today}
\begin{abstract}
We introduce a new model for the generation of random satisfiability
problems. It is an extension of the 
hyper-SAT model of Ricci-Tersenghi, Weigt and Zecchina,
which is a variant of the famous $K$-SAT model: it is extended to 
$q$-state variables  and relates to a different choice of the statistical
ensemble.

The model has an exactly solvable statistic:
the critical exponents and scaling
functions of the SAT/UNSAT transition are calculable 
at zero temperature, with no need of replicas, 
also with exact finite-size corrections.

We also introduce an exact duality of the model, and
show an analogy of 
thermodynamic properties with the
Random Energy Model of disordered spin systems theory.
Relations with Error-Correcting Codes are also discussed.
\end{abstract}

\PACS

\noindent
{\it keywords:} Random satisfiability,
Random Energy Model,
error-correcting codes,
hyper-graphs,
random linear systems.


\section{Introduction}

\subsection{Motivations}

Complexity theory concerns the classification of combinatorial optimization
problems, according to the computational cost required for
their solution~\cite{NPC,goldreich,wilf}.

A central role is occupied by the class of hard problems
named NP (Non-deterministic Polynomial time). For problems in this class, a
potential solution can be checked in polynomial time,
whereas finding one solution may require, in the worst case,
a time growing exponentially with system size. 

Often, the crude worst-case analysis of ordinary complexity theory hides the
mechanism which leads to NP complexity. A statistical mechanical analysis of
the ensemble of possible instances of the problem 
(we could say a ``typical case'' analysis,
eventually dependent on some macroscopic parameter of the theory) would have
the ambitious challenge of deepening the understanding of computational
complexity, and improving the algorithm design 
for real world applications.

For example, in a variety of cases, the problem admits 
a space of parameters, and a certain order parameter exists, showing a
critical behaviour. In some of these cases, a widespread conjecture is
that the computationally hard instances appear with a significant
probability only when generated near ``phase boundaries'' of the ensemble,
with respect to this phase transition.

An exemplar case of this phenomenon 
is the random Boolean satisfiability problem
($K$-SAT~\cite{satrev1,satrev2,chayes}).
In this case, a pecular transition, called {\em SAT/UNSAT transition}, occurs:
when the density of clauses is higher than a certain critical value, the system
is highly constrained, and the average probability of having a solution goes
to zero in the thermodynamic limit. Otherwise,
the system is slightly constrained, and the average probability of 
having a solution goes to one in the thermodynamic limit.
Hard-to-solve instances can be found only in a narrow region
on the density axes, in correspondence of the critical density.
This fact has a clear interpretation:
heuristic algorithms for randomly searching a solution are efficient when the
density is low enough, while algorithms for randomly searching a contradiction
are efficient when the density is high enough.
At the phase boundary, however,
both the average probability of being satisfable or unsatisfiable are order 1,
then heuristic (incomplete) search algorithms have no way to disentangle,
in a given finite time, the unsatisfiable instances, from those which
are simply very hard to solve.
So, in this region
there appears an exponential critical slowing down which makes the
search inefficient for any practical purpose.

A sub-class of the NP class exists, called NP-complete, 
such that, for each problem in this sub-class,
no other NP problem is more than a polynomial factor harder.
Informally, a
problem is NP-complete if answers can be verified quickly (NP requirement), 
and a quick algorithm to solve this problem can be used to solve 
all other NP problems quickly (completeness requirement).
NP-complete problems are particularly important, as, in some sense, they
manifest all the potential difficulties of any NP problem, and are a sort of
``language'' for properly describing every possible problem in NP class.
The $K$-SAT problem 
is a NP-complete problem for $K\ge 3$. For its simplicity in formulation, and
versatily in encoding other NP problems, it has acquired a paradigmatic role
for the study of NP complexity. It has been studied both directly 
(see e.g.~\cite{ach1,ach2}, or, for a review,
\cite{satrev2} and references therein), or by the introduction of
more ``handable'' variants which reproduces some of the peculiarity of $K$-SAT
problems. One step in this direction is the introduction of the {\em hyper-SAT}
model by F.~Ricci-Tersenghi, M.~Weight and R.~Zecchina (see \cite{rwz,lrz}).

\subsection{Presentation of the material}

In this paper, we shall discuss a very simple and exactly solvable
model for the generation of random combinatorial problems. The model
has deep analogies with the hyper-SAT model,
and inherits from it some important characteristics:
\begin{itemize}
\item it is hard to handle by local search
methods in the whole region of the parameter space;
\item the problem may be 
solved in polynomial time by a simple global method
(Gauss reduction) and
therefore belongs to the class P;
\item in a variant (unfrustrated model) one
solution may be superimposed by construction.
\end{itemize}
The most important feature of our model
is the fact that
also the statistical properties of the model 
are exactly known, at finite and infinite size, thus yielding to
a wide amount of analytical results.
For example, we obtain the exact expression for the probability distribution of
the rank of a $M \times N$ random matrix in $GF(q)$, result which was
previously known only in the limit $N \rightarrow \infty$ at constant
$M-N$.
This result contains all the information about the average
``satisfiability'' probability, and much more. 
The methods used in the paper can be extended in order to calculate other
interesting quantities.

Furthermore, the simple underlying geometry of the model allow for a
clarifying graph-theoretical interpretation of the problem, which leads to the
formulation of an exact duality of the system.

\vbox{The paper is organized as follows:
\begin{enumerate}
\item In this introductory section we shortly
describe some problems related to our model: random linear systems, the
$K$-SAT model and the hyper-SAT model.
\item In the second section we discuss of
some related fields of application, mainly Error Correcting Codes Theory and
Hypergraphs Theory.
\item In the third section we introduce the model, and describe 
its main properties, with particular attention to the duality and relation
deriving from it.
\item In the fourth section we exactly calculate the free energy
distribution function at zero temperature, 
solving a realization of the ``classical 
moment problem''.
\item In the fifth section we rederive the results of the previous
section, by mean of a direct analysis of the Gauss algorithm
which allows for a depiction
of the structure underlying the ensemble of instances.
\item In the sixth section 
we derive the conclusions of our work, and
  anticipate the directions of our forthcoming research.
\end{enumerate}}

\subsection{A prolog: systems of random linear equations in $GF(q)$}

We will start with a small prolog of elementary linear algebra, as it will
provide a nice setting for our model.

Systems of linear equations, beyond the ordinary theory on 
$\mathbb{R}$ or $\mathbb{C}$,
can be studied on more generic structures.
For example, in the textbook \cite{kolchin}, following the original work of 
Kovalenko and collaborators \cite{kova1, kova2}, a certain number of questions
about systems of linear equations on the Galois field $GF(q)$ is answered.

The Galois field $GF(q)$, with $q$ a prime number,
is a finite field whose elements are the
integers in $\mathbb{Z}_q$ (i.e.~the integers from 0 to $q-1$), and sum
and product are defined modulo $q$:
\begin{subequations}
\begin{align}
(x_1 + x_2)_{GF(q)} &\equiv x_1 + x_2 \modulo{q} \ef{;}
&
(x_1 + x_2)_{GF(q)} &\in \mathbb{Z}_q \ef{;}
\\
(x_1 \cdot x_2)_{GF(q)} &\equiv x_1 \cdot x_2 \modulo{q} \ef{;}
&
(x_1 \cdot x_2)_{GF(q)} &\in \mathbb{Z}_q \ef{.}
\end{align}
\end{subequations}
The prototype of a linear system $A \vett{x} = \vett{v}$, or of a
homogeneous linear system $A \vett{x} = \vett{0}$, in $GF(q)$, 
when the above definitions
are assumed, maintains the same formal structure. It also preserves most of
the fundamental results of linear algebra, like the existence of a Gauss
algorithm for triangulating a matrix, or the fact that the
rank plus the dimension of the kernel is equal to the number of rows of the
matrix.

Statistical mechanics plays a role when we consider averages over a certain
ensemble of matrices and vectors, in a sort of ``thermodynamic limit'' on
their dimension.

For example, given two integers $N$ and $M$, and a prime number $q$,
a good statistical ensemble is the set of the couples $(A,\vett{v})$, with
$A$ an $M \times N$ matrix, and $\vett{v}$ an $M$-dimensional
vector, with elements randomly chosen in $\mathbb{Z}_q$.
A statistical mechanic would calculate, for a certain fixed value of
$\gamma=N/M$, in the limit $N \rightarrow \infty$,
the average satisfiability of this problem, i.e., for a random
``extraction'' of
an instance $(A,\vett{v})$ from the ensemble, 
which is the probability that the linear system in $GF(q)$, 
$A \vett{x} = \vett{v}$, 
has at least one solution. In more detail, we could calculate 
which is the probability
distribution for the dimension of the kernel of the linear application
$A$, in the limit of infinite $N$ and $M$, with $N-M$ fixed.

This last point is exactly the main question answered by the Russian school in
\cite{kolchin}, which will be rede rived as a preliminary result in this paper.

\subsection{Nomenclature}

In the framework of statistical mechanics, it is frequent to work with spin
variables $\sigma = \pm 1$, while
in the framework of random satisfiability, it is customary to work with
Boolean variables $b \in \{ \true, \false \}$.
We will often switch between the two formulations, with the 
table of correspondences
\begin{equation}
\label{tavcorr}
\begin{array}{|c|c|}
\hline
        b      & \sigma \\
\hline
        \true  & 1      \\
        \false & -1     \\
\hline 
\end{array}
\end{equation}
So, we will consider a set of $N$ spin variables
$\sigma_j$, with $j=1, \ldots, N$, or, alternatively, 
a set of $N$ Boolean variables $b_j$. Each of the $2^N$ 
choices of the variables, denoted by $\sigma$, or by $\vett{b}$, will be
called a \emph{configuration}.

A random satisfiability model 
is defined in terms of an ensemble of random \emph{instances},
that is a collection of $M = N/\gamma $ 
Boolean conditions on the variables (called \emph{clauses}), randomly
chosen from a certain set.
This set of the possible clauses will primarily distinguish the 
different models,
so it will be discussed further, case by case.

Given a certain instance, if we have at least one
configuration which simultaneously
satisfies all the clauses\footnote{It will be 
``one non-trivial configuration'' in
  the model with a superimposed solution.}, 
the instance is a \emph{satisfiable} (SAT)
one, else it is an \emph{unsatisfiable} (UNSAT) one.

We will introduce a fictitious
Hamiltonian for each instance. Our choice is to assign to a configuration 
an energy equal to the number of unsatisfied clauses.
The low temperature limit of the
partition function calculated with this Hamiltonian 
is the number of solutions for a certain instance.

We will adopt the graphical convention for the logical operators 
$\band$ (and), $\bor$ (or) and $\bxor$ (xor).

\subsection{The $K$-SAT model}

In this model, given an integer $K$, the set of clauses $\{ C \}$ is
the set of all the $K$-uples of variable indices $n_1, \ldots
,n_K$, with a Boolean assignment $s_j$ attached to each index.
An instance in the ensemble is a subset 
$\mathcal{I}$ 
of this set of 
clauses, with
cardinality $M$, and with no repetitions admitted.

A certain configuration will be a solution if
\begin{equation}
\vett{b} \ \textrm{solut.}
\quad
\Longleftrightarrow
\quad
\underset{C_i \in 
\mathcal{I}
}{\bigwedge} 
\left( \underset{j\leq K}{\bigvee} 
\left( s_j^{(i)} \bxor \, b_{n_j^{(i)}} \right) \right)= \true
\ef{.}
\end{equation}
If we call $\sigma_j$ the spin variables associated to the Boolean variables 
$b_j$, and
$\tau_j^{(i)}$ the spin value associated to the Boolean assignments
$s_j^{(i)}$, both following the table (\ref{tavcorr}),
a possible spin Hamiltonian which counts the unsatisfied clauses is
\begin{equation}
\ham_
{\mathcal{I}}
=\frac{1}{2^K} \sum_{C_i \in \mathcal{I}}
\prod_{j=1}^{K}
\left( 1 - \tau_j^{(i)} \sigma_{n_j^{(i)}}
\right)
\ef{.}
\end{equation}

Interpreted as a spin model Hamiltonian, 
the system has random diluted 
quenched
couplings, with a complicated pattern of interactions:
for each $K$-uple of variables, if we have a $K$-spin coupling between them,
then we have also couplings between all the subsets of the $K$ variables, with
a choice of signs
causing the $2^K$ levels of the
$K$-variables $1$-clause
subsystem to split in a $2^K-1$ multiplet of energy $-1$, and 
one level of energy
$2^K-1$.

\subsection{The hyper-SAT model}

This model,
introduced by Ricci-Tersenghi, Weigt and Zecchina in \cite{rwz} (see also
\cite{lrz}) as a variant of the $K$-SAT with a more clear spin structure, 
is defined as follows.

Given an integer $K$, 
the set of clauses $\{ C \}$ is the set of all the $K$-uples 
of variable indices $n_1, \ldots
,n_K$, with a single Boolean assignment $s$ attached. 
An instance in the ensemble is a subset $\mathcal{I}$ of this new 
set of clauses, with
cardinality $M$, and no repetitions admitted.

A certain configuration will be a solution if
\begin{equation}
\vett{b} \ \textrm{solut.}
\quad
\Longleftrightarrow
\quad
\underset{C_i \in \mathcal{I}}{\bigwedge} 
\left( 
\Bigg( \underset{ \substack{j\leq K} }
{\dot{\bigvee}} b_{n_j^{(i)}} 
\Bigg)
\, \bxor \; s^{(i)}
\right)=\true
\ef{.}
\end{equation}
If we call
$\tau^{(i)}$ the spin value associated to the Boolean assignments
$s^{(i)}$, again following the table (\ref{tavcorr}),
a possible spin Hamiltonian which counts the unsatisfied clauses is
\begin{equation}
\label{9435}
\ham_{\mathcal{I}}=
\frac{1}{2}
\sum_{C_i \in \mathcal{I}}
\Big( 1 - \tau^{(i)}
\prod_{j=1}^{K} \sigma_{n_j^{(i)}}
\Big)
\ef{.}
\end{equation}
This ensemble of Hamiltonians has the form of a 
diluted mean field $K$-spin disordered system, that is a
spin model with $N$ variables and $M$ quenched $K$-spin couplings, 
randomly chosen between all the $K$-uples
of variables, and with values randomly chosen in $\{ -1, 1 \}$.

A variant of the problem, named ``unfrustrated'' model, 
is the restriction of the clauses
set only to the clauses with $s=\true$ 
(that is, a ferromagnetic diluted $K$-spin
system). In this case we will have the
obvious ``ferromagnetic'' solution $b_j= \true $ (or $\sigma_i=+1$) 
for each instance. 

\subsection{The hyper-SAT model as a random linear systems}

The hyper-SAT model can be reinterpreted in the framework of random linear
systems we introduced in the prolog. Given a hyper-SAT instance
with $N$ variables and $M$ clauses, we can introduce an $M \times N$ 
matrix $\grid_{ij}$
in the field $GF(2)$, such that $\grid_{ij}=1$ if the variable $b_j$ appears
in the $i$-th clause, and $\grid_{ij}=0$ elsewhere.
With the expanded table of correspondences
\begin{equation}
\label{tavcorr2}
\begin{array}{|c|c|c|}
\hline
        b      & \sigma & \ x \ \\
\hline
        \true  & 1      & 0 \\
        \false & -1     & 1 \\
\hline 
\end{array}
\end{equation}
we have the composition correspondences
\[
b_1 \bxor b_2 = \sigma_1 \sigma_2 = (x_1 + x_2)_{GF(2)}
\ef{;}
\]
so, for each clause $C_i$,
\[
\dot{\bigvee_{j | J_{ij}=1}} b_j =
\prod_j \sigma_j^{J_{ij}} =
\left( \sum_j J_{ij} x_j \right)_{GF(2)}
\ef{.}
\]
So, at the end, we can analyze the problem in the more handable 
form $\grid \vett{x} = \vett{v}$
(frustrated model), or $\grid \vett{x} = \vett{0}$
(unfrustrated model).

The shift to $K$-spin models will be given by the
correspondence (\ref{tavcorr2}), so the Hamiltonian in the spin formulation
will be
\begin{equation}
\ham_{\grid, \vett{v}}(\sigma)= \frac{1}{2} \sum_i \Big( 1- (-1)^{v_i}
\prod_j \sigma_j^{\grid_{ij}} \Big) \ef{.}
\end{equation}
while in the linear system formulation will be
\begin{equation}
\label{1764}
\ham_{\grid, \vett{v}}(\vett{x})= \sum_i 
\Big( 1 - \delta_2 
\Big(\sum_j \grid_{ij} x_j - v_i \Big) \Big) \ef{.}
\end{equation}
with $\delta_2(n)=1$ if $n$ is even, and $0$ if $n$ is odd.
As in the following we will generalize the system to $q$-state variables,
we will introduce a $\delta_q$ function with the definition: 
\begin{equation}
\label{deltaq}
\delta_q (n) = \left\{
\begin{array}{lr}
1 & n 
\ \, \textrm{\makebox[-0.35\width][l]{\phantom{/}}} {\equiv} \ 
0 \modulo{q} \\
0 & n 
\ \, \textrm{\makebox[-0.35\width][l]{/}} {\equiv} \ 
0 \modulo{q} 
\end{array}
\right.
\ef{.}
\end{equation}
Beyond the correspondence between a specific hyper-SAT instance and a specific
random linear system, we remark that the choice of the ensemble is crucial
for the statistical properties of the system. 
The statistical
ensemble of the hyper-SAT is very different to the random ensemble we
suggested as an example in the prolog, as in the hyper-SAT we have 
exactly $K$ non zero terms
in each row of $\grid$.

\section{Random satisfiability, graph theory and Random Linear Codes.}

\subsection{Linear codes in information theory}
\label{subsec538}

A main problem in information theory consists in signal reconstruction when a
sequence of bits is transmitted through a
``noisy'' channel, that is a channel where
transmission errors can occur with a certain probability.
Just to fix the ideas, we will
concentrate on the so called {\em symmetric noise channel}, in which each
bit has a probability $(1-p)$ to be transmitted unaltered, and a
probability $p$ to be transmitted as its negate:
\[
\begin{array}{|c|cc|}
\cline{2-3}
\multicolumn{1}{c|}{}  & 0 & 1
\\
\hline
0 & 1-p & p 
\\
1 & p & 1-p
\\
\hline
\end{array}
\]
When $p=0$ no error can occur, and transmission is for granted. When $p=1/2$
the output signal is totally decorrelated from the input signal, and no
inference is possible even in principle. In the intermediate cases, the task
is to use the information transmitted through the channel, in the form of 
the partial correlation between input and output signals, to obtain the safe
transmission of a shorter signal.

This problem can be easily generalized to a generic ``alphabet'', with 
characters in $\mathbb{Z}_q$ (i.e., instead of having bits in $\{0,1 \}$, 
we have variables in $\{0, 1, \ldots, q-1 \}$).
A generalization of the symmetric noise channel defined above could be as
follows: each
character has a probability $(1-p)$ to be transmitted unaltered, and a
probability $\frac{p}{q-1}$ to be converted in each of the other
characters. For example, when $q=3$ we would have
\[
\begin{array}{|c|ccc|}
\cline{2-4}
\multicolumn{1}{c|}{}  & 0 & 1 & 2
\\
\hline
0 & 1-p & p/2 & p/2 
\\
1 & p/2 & 1-p & p/2
\\
2 & p/2 & p/2 & 1-p
\\
\hline
\end{array}
\]
In this case, the totally decorrelated case occurs for $p=\frac{q-1}{q}$.

A \emph{code} is an algorithm which fulfills the task above,
replacing the interesting signal with a redundant but error-robust 
encoding of it. The ratio between the length of the original and of the
encoded signals is called the \emph{rate} of the code.
A critical rate exists such that, in the limit of infinite length of the
message, the decoding procedure almost always succeeds in reconstructing
the original signal from the noise-distorted transmitted stream.
In most cases, 
the probability of success as a function of the rate 
jumps to one in an abrupt way, typical
of critical phenomena (reconstructable-unreconstructable transition).

The quality of a code
is mainly given by two parameters: the critical 
\emph{rate} at given error probability
and the computational complexity class of the
encoding and decoding algorithms.

A family of codes, called \emph{linear codes} \cite{roman}, is closely related
to random satisfiability problems and certain spin glass models in statistical
mechanics \cite{sourlas98}, and a deep insight into the
reconstructable-unreconstructable transition can be derived from the study of
the phase diagrams of the related spin glass models \cite{montan01}. A well-known example of linear code, called
\emph{Varshamov's linear code}, or \emph{RLC} (\emph{Random Linear Code})
\cite{rlc1, sudan} 
has been the first discovered 
code to reach the \emph{Gilbert-Varshamov bound} on the rate of the code.
The Varshamov's code is closely related to our model.

Suppose we have a random linear system $\grid \vett{x} = \vett{v}$ 
in $GF(q)$, with $M$ equations, in $N>M$ variables, and 
suppose the kernel of the
linear application $\grid$ has the minimum possible dimension 
$L=N-M$. This linear system can be used for the error correcting
transmission of a message of length $L$, on an alphabet of $q$
characters.
First we introduce some definitions:
\begin{enumerate}
\item The possible original messages are the vectors $\vett{y}$ of the linear
  space ${\mathbb{Z}_q}^{L}$.
\item The encoding map $\mathcal{E}
\big( {\mathbb{Z}_q}^{L} \rightarrow {\mathbb{Z}_q}^{N} \big)$, which
encodes the original $L$-bits message in the redundant $N$-bits 
codeword,  
is defined with the aid of the linear application
$\grid$. If we choose a basis for the kernel
\begin{equation}
\kker (\grid) = \sspan (\vett{s}_{(1)}, \ldots \vett{s}_{(L)} ) 
\ef{,}
\end{equation}
the map will be defined by
\begin{equation}
\mathcal{E} :
\quad
\vett{y} \rightarrow \vett{x}(\vett{y}) = \sum_{i=1}^{L} y_i
\vett{s}_{(i)} \ef{.}
\end{equation}
\item A distance on the codewords space ${\mathbb{Z}_q}^N$ will be introduced:
\begin{equation}
d \left( \vett{x}, \vett{x}' \right) =
\sum_{j=1}^{N} \left[ 1 - \delta (x_j - x'_j) \right] \ef{.}
\end{equation}
\end{enumerate}
Then, for the transmission of a certain message, we follow the procedure:
\begin{enumerate}
\item The original message is a vector 
$\vett{y}_{\textrm{mes}} \in {\mathbb{Z}_q}^{L}$.
\item We use the map $\mathcal{E}$ for the encoding. The original codeword 
is $\vett{x}_{\textrm{mes}} = \mathcal{E} (\vett{y}_{\textrm{mes}})$.
\item The codeword is transmitted through the noisy channel. The transmitted
  codeword will be a vector $\vett{x}_{\textrm{trans}}$. The probability
  distribution for the transmitted codeword depends on the kind of noise we
  consider. In the case of symmetric noise we have
\begin{equation}
p(\vett{x}) = (1-p)^N 
\left( \frac{p}{(1-p) (q-1)} \right)^{d(\vett{x}, \vett{x}_{\textrm{mes}})}
\ef{.}
\end{equation}
\item for the reconstruction of the signal could be used, for example, the
  Minimum Distance Decoding criterium, that is, if we call
$\vett{x}_{\textrm{rec}}$ the minimum of the function 
$d(\vett{x}, \vett{x}_{\textrm{trans}})$ in the image of the encoding map
$\mathcal{E} \big( {\mathbb{Z}_q}^{L} \big)$, 
 we choose as the reconstructed message $\vett{y}_{\textrm{rec}}$
the vector 
$\vett{y}_{\textrm{rec}} = \mathcal{E}^{-1} 
\left( \vett{x}_{\textrm{rec}} \right)$.
\end{enumerate}
The interesting question is, in the ``thermodynamic'' 
limit of long signals
$N \rightarrow \infty$, with the {\em rate}
$R = L / N$ fixed,
which is the probability that the reconstructed message
$\vett{y}_{\textrm{rec}}$
is identical to the original message
$\vett{y}_{\textrm{mes}}$,
as a function of the noise parameter $p$ and of the rate
$R$.

A statistical study \cite{montan01} shows the tricky fact that 
this probability tends to one if $R < R_c(p)$, and it tends to zero
if $R > R_c(p)$, and that the critical line $R_c(p)$ is
exactly the Shannon bound for the capacity of a symmetric noise channel 
\cite{roman, sudan}.

\subsection{Random graphs and statistics}

Another mathematical framework in which this kind of satisfiability problems is
well depicted is hypergraph theory. In these sections we will shortly 
introduce some notation which will be useful in the following, first 
starting with
ordinary graph theory, then jumping to hypergraphs (for an introductory
bibliography, see \cite{bollobas, berge}).

Given
\begin{displaymath}
V=\{\punto_1, \ldots , \punto_N \}
\end{displaymath} 
a finite set of cardinality $N$, called the set of
``vertices'',
we define the set of all the non-ordered 
couples of vertices
in $V$
\begin{displaymath}
\mathcal{E}= \{ (\punto_{n_1},\punto_{n_2}) \} \qquad n_1 \neq n_2
\end{displaymath}
as the set of ``edges''. 
For each subset 
$E \subseteq \mathcal{E}$ of cardinality $M$, the datum of the set of vertices
and of this set is
called the \emph{graph} $\Gamma(V,E)$, with $N$ vertices and $M$ edges.

Given a
vertex $\punto_i$, its \emph{coordination} (or \emph{degree}) is defined
as the number of edges $E_j \in E$ such that $\punto_i \in E_j$.
A \emph{path} between the vertices $\punto_a$ and $\punto_b$ is a sequence of
vertices and edges in $E$
\begin{displaymath}
(\punto_{a}, E_{1}, \punto_1, E_{2}, \punto_2, \cdots,
E_{L}, \punto_{b})
\end{displaymath}
such that $\punto_i \in E_{i}, E_{i+1}$;
the number of edges $L$ is called its \emph{length}. A graph is said
\emph{connected} if, for each couple of vertices, exists a path which 
joins them.

A non-trivial closed path is usually called a \emph{loop}. More generically, we
will adopt the ``physical'' convention of using the term loop for any subset
of the set of edges $E_L \subseteq E$ such that the graph, restricted to this
set $\Gamma(V,E_L)$, has only vertices of even (eventually zero) coordination.

In the limit of large $N$ and $M$, we can introduce an ensemble of graphs,
giving a probability measure over the possible sets of edges, and we can
calculate statistical averages of interesting quantities. For example, an
ensemble could be the one in which only connected graphs with $M$ 
edges are admitted, and no repetitions of edges are allowed.

\subsection{Hypergraphs and $K$-SAT problems}

A generalization of graph theory concerns objects in which the set of possible
edges $\mathcal{E}$ is replaced by the set of all non-ordered $K$-uples of
vertices (typically $K \ge 3$), called \emph{hyper-edges}:
\begin{displaymath}
\mathcal{E}_K= \{ (\punto_{n_1},\punto_{n_2}, \cdots, \punto_{n_K}) \}
\qquad n_1 \neq n_2 \neq \cdots \neq n_K \ef{.}
\end{displaymath}
The datum of the set of vertices and of a subset of admitted hyper-edges will
be called a \emph{hypergraph}.

The notions of coordination, 
path, connectivity, and loop, as we defined them in the previous
section, immediately generalize to this case. In particular, the notion of
\emph{hyper-loop}, as a subset $E_L \subseteq E$ 
such that the relative hypergraph 
$\Gamma(V,E_L)$ has only vertices of even coordination, will be crucial 
in the following.

The ensemble of connected hypergraphs with $M$ hyper-edges, with
repetitions allowed, is the proper graphical scheme for describing the
topological features of ordinary $K$-SAT problems, and also of the hyper-SAT,
although, as we already mentioned in the previous section, we should also
imagine some Boolean assignments 
attached to the hyper-edges (one for each vertex
of each hyper-edge in the case of $K$-SAT, just one per hyper-edge in the case
of hyper-SAT, and no assignments at all in the case of unfrustrated
hyper-SAT\footnote{This immediate correspondence with the hypergraph
  interpretation is the reason for the name.}).

Another possible variant could be the one in which hyper-edges
with any number of vertices are admitted
\begin{displaymath}
\mathcal{E}_{\forall}= \bigcup_{K=0}^{N} \mathcal{E}_{K} \ef{.}
\end{displaymath}
The ensemble of all possible hypergraphs of $M$ hyper-edges in
$\mathcal{E}_{\forall}$, with 
homogeneous measure, and no
repetitions allowed, 
is the one which describes the model we will
define in the next chapter.

\subsection{Group structure of the hyper-SAT and 
the role of hyperloops in satisfiability}
\label{subsec7452}

In the hyper-SAT model the configurations shows a 
${\mathbb{Z}_2}^N$ group structure. 
The group operation is the
sum term-by-term in the linear system formulation, or
the product term-by-term in the spin formulation, that is
\begin{align}
\label{prod}
x_i^{(a b)} & \equiv x_i^{(a)} + x_i^{(b)} \modulo{2} \ef{;}
&
\sigma_i^{(a b)} & = \sigma_i^{(a)}
\sigma_i^{(b)}
\ef{.}
\end{align}
In the unfrustrated model, the set of solutions is a
subgroup ${\mathbb{Z}_2}^{g}$, with $g$ 
generators, of the configuration group.

The fact that in the
unfrustrated hyper-SAT
the set of solutions is a subgroup
is an immediate consequence of the fact that the kernel of a linear
application in $GF(2)$
is a subspace ${\mathbb{Z}_2}^{g}$ of the vector space
${\mathbb{Z}_2}^{N}$.

Let's go back to the formulation of a 
hyper-SAT instance in
terms of a system of linear equations in $GF(2)$:
\[
\sum_{j=1}^{K} x_{n_j^{(i)}}=v_i 
\ef{.}
\]
We can graphically identify each variable with a vertex, and each clause with
a hyper-edge joining the vertices which appears in it.

Each subset of equations in the system can be identified with the 
sub-hypergraph (in the following just hypergraph) 
of the original one, obtained by restriction to the
correspondent hyper-edges. The sum of the equations in
the subset will be called the \emph{equation associated to the hypergraph}.

Note that also hypergraphs have a group structure
${\mathbb{Z}_2}^{M}$: given two hypergraphs, the group operation is 
the disjoint
union of the two hypergraphs, from the graphical point of view,
and the sum of the two corresponding
equations, from the algebraic point of view.
The identity is the empty set.

The set of hyperloops is a subgroup
${\mathbb{Z}_2}^{g'}$ of the hypergraph group.
When the subgraph is a hyperloop, 
the associated equation is a relation
involving no variables:
in the unfrustrated case, it is just a trivial identity; 
in the frustrated case, it is
a configuration-independent
characteristic of the instance which could eventually determine an 
\emph{a priori} unsatisfiability.
Because of randomness in the choice of the $s_{(j)}$ for the
instance, this will happen with $1/2$ probability for each linearly
independent hyperloop. The number of linearly independent elements in a group 
${\mathbb{Z}_2}^{g'}$ is just $g'$.

So the sentence in \cite{rwz} which explains the role of hyperloops in the
frustrated model:
\myquote{As soon as one 
hyperloop arises in the hypergraph, half the formulas
become unsatisfiable...}
can also be
interpreted as 
``As soon the average number of generators for the hyperloop group, 
$g'$, increases of one unity, the average number of solutions would 
decrease
of a factor one half
(if we were working with the frustrated model), 
so the average number of generators of the group
of solutions, $g$, decreases of one unity''. 

We can see that an intriguing
symmetry of roles between solutions and
hyperloops seems to arise.
This symmetry of roles 
is more clear in the linear system formulation, as we have
that a homogeneous solution is a solution of 
$\grid \vett{x}= \vett{0}$, while a hyperloop is a solution of
$\grid^T \vett{y}= \vett{0}$.
In our model, the analytical solvability will provide a clarifying
reinterpretation of these heuristic arguments.

\section{The model}

One difficulty on an exact analytic approach to the
hyper-SAT is that the ensemble of formulas is
\emph{not} symmetric: for an admitted clause, the number of 1 in each
row must be exactly $K$, while there is not any similar constraint 
on the columns.
This fact introduces difficult-to-handle 
correlations on the values of the entries in the grid formulation.

In this section we introduce a model defined on a larger ensemble, which 
cancels correlations between the entry of the matrix, and
fully exploits the symmetry of roles between solutions and hyperloops we
suggested in section \ref{subsec7452}.

\subsection{Description of the model}

The model  we are going to define is a random satisfiability problem. So
it will be defined by an ensemble of {\em instances}, with a probability
measure on it (the possible
realizations of the problem, with associated the probability that such a
realization occurs), an ensemble of 
{\em configurations} (the potential solutions), and a notion of 
{\em satisfiability}.

Given the integer parameter $q$, which must be a prime number,
and the ``noise'' parameter $p \in [0,1]$
(as it will have a role similar to the one of the noise parameter of Error
Correcting Codes, see section \ref{subsec538}),
the ensemble of the instances is the
product set of
the set of all the $M \times N$ matrices
$\grid_{ij}$, filled with integers in $\mathbb{Z}_q$, each matrix being
equiprobable, and the $M$-dimensional vector space ${\mathbb{Z}_q}^M$, where
the probability for a vector $\vett{v}$ to be extracted is dumped accordingly
to the number of its non-zero entries.
More precisely
\[
\mu(J,\vett{v})= q^{-MN} \cdot \left( 1-p \right)^M
\left( \frac{p}{(1-p)(q-1)} \right)^{\sum_{i=1}^{M} (1-\delta_q(v_i))}
\ef{;}
\]
where we used the short notation of equation (\ref{deltaq}); the parameter $p$
has been chosen to reproduce the average fraction of non-zero elements in
$\vett{v}$. The parameter $\gamma=N/M$ will determine the average solvability
of the problem: large values of $\gamma$ corresponds to weakly constrained
systems, and a typical instance will be solvable, while, on the contrary,
small values of $\gamma$ correspond to highly constrained systems, and a
typical instance will be unsolvable (or will have only the 
superimposed solution in the unfrustrated case, see below).

The ensemble of the configurations is the $N$-dimensional vector space 
${\mathbb{Z}_q}^N$. A certain configuration $\vett{x}$ is said to be a
solution if it solves the linear system
\[
J \vett{x}= \vett{v}
\ef{.}
\]
The particular case of $p=0$ is called {\em unfrustrated}, as we have
a superimposed ``ferromagnetic'' solution $\vett{x}$ for each instance
$\grid$.
When $p \neq 0$ we will deal with the {\em frustrated} version.
When $p$ reaches the value $p=(q-1)/q$,
the model will be called {\em totally frustrated}, as the
probability measure is flat also for the inhomogeneous terms $\vett{v}$.

A particular realization of the problem ({\em instance}) will be said 
{\em satisfiable} (SAT) if we have at least one solution, in the frustrated
model, and at least one more solution, after the superimposed one, in the
unfrustrated model. Elsewhere, the instance will be said to be 
{\em unsatisfiable} (UNSAT).

Although in this part we only study the average solvability of the problem, 
it is customary of a statistical
mechanical approach to introduce a fictitious Hamiltonian, such that the
solutions correspond to the states with zero energy, and all the other
configurations have a rigorously positive energy, study the thermodynamic of
this new system, and recover the original model in the zero temperature limit.

Obviously, there is a large arbitrarity in the choice of the Hamiltonian, and
we will analyze various possibilities in the third part of the paper. Anyway,
just to fix the ideas, we will concentrate on the most simple choice, the one
in which the Hamiltonian just ``counts'' the number of violated clauses:
\begin{equation}
\label{hamiltx}
\ham_{\grid}(\vett{x})= \sum_{i=1}^{M}
\left( 1- \delta_q 
\Big( \sum_{j=1}^{N} \grid_{ij} x_j -v_i \Big) \right)
\ef{,}
\end{equation}
A more familiar spin reformulation is
\begin{equation}
\label{8637}
\ham_{\grid}(\sigma) =
\frac{1}{2}
\sum_{i=1}^{M}
\left( 1- 
\delta \Big( s_i \prod_{j=1}^{N} \sigma_j^{\grid_{ij}} -1 \Big) \right)
\ef{.}
\end{equation}
where the correspondence is
\[
\sigma_j=e^{\frac{2 \pi i}{q} x_j}
\ef{;}
\qquad 
s_i =e^{\frac{2 \pi i}{q} v_i}
\ef{.}
\]
In the
homogeneous case every $s_i=1$.

This model shares with the Random Energy Model \cite{MPV,ruelle,RE1,RE2,RE3},
the property that the
energy of different states are (almost always) uncorrelated. 
More precisely, it is true that, if $(E, E_1, \ldots , E_k)$ are the energy
respectively of the states $(\sigma, \sigma^{(1)}, \ldots, \sigma^{(k)})$,
calculated with a certain instance, for
each couple of functions $f(x)$, $g(x_1, \ldots, x_k)$ we have
\[
\reval{f(E) \, g(E_1, \ldots, E_k)}= 
\reval{f(E)} \cdot \reval{g(E_1, \ldots, E_k)}
\]
if $\sigma \notin \sspan ( \sigma^{(1)}, \ldots , \sigma^{(k)} )$, and the
fraction of states for which this last condition is verified
is about $1-q^{-N+k}$.

\subsection{The underlying group structure}
\label{subsec6598}

As we anticipated in section \ref{subsec7452},
a nice feature of our model, which is common also to the hyper-SAT model and
to the random linear system problem, is that the configurations shows a 
$G_{\sigma} = {\mathbb{Z}_q}^N$ group structure. 

The group operation in the linear system formulation is the
sum term-by-term of the configuration entries:
\begin{equation}
\label{prodbis}
x_i^{(a b)} \equiv x_i^{(a)} + x_i^{(b)} \modulo{q} \ef{;}
\ef{.}
\end{equation}
In the unfrustrated model, the set of solutions is a
subgroup $G_S = {\mathbb{Z}_q}^{g}$, with $g$ 
generators, of the configuration group.

In the frustrated model, given an instance $(\grid, \vett{v})$,
we define the \emph{associated homogeneous problem}
the one in which we search for solutions of the homogeneous 
system $\grid \vett{x}=\vett{0}$
(that is, the problem in which
we turn all the Boolean assignments $s_{(j)}$ to $\true$)
As we said above, the set of solution of this homogeneous problem is a
subgroup 
$G_S = {\mathbb{Z}_q}^{g}$.
If we have at least one
solution $\widehat{\vett{x}}$ 
for the original problem, the set of
solutions of the original problem is the lateral class 
$\widehat{\vett{x}} + G_S$.

These statements have a clarifying correspondence
in the random linear system on $GF(q)$ framework:
the space of admitted configurations $\vett{x}$ is the discrete torus
${\mathbb{Z}_q}^{N}$, with, as a group operation, 
the sum of discrete vectors
quotiented by periodicity.
The fact that in the
unfrustrated hyper-SAT
the set of solutions is a subgroup,
and in the frustrated case it is a lateral class,
is in correspondence with the fact that the set of solutions of a linear
system on $GF(q)$
is a subspace ${\mathbb{Z}_q}^{g}$ of the vector space
${\mathbb{Z}_q}^{N}$ in the homogeneous case, 
and an affine subspace in the inhomogeneous case.

The model naturally shows also a gauge invariance.
As the Hamiltonian is of the form
\[
\ham_{\grid, \vett{v}}(\vett{x}) = \ham(\grid \vett{x}-\vett{v})
= \ham(\vett{w})
\ef{,}
\]
it will remain unchanged for transformations which keep unchanged the vector
$\vett{w}$, that are the changes of basis and the translations on the torus of
configurations ${\mathbb{Z}_q}^{N}$. The more general gauge transformation
$\mathcal{G}$ of
this kind is
\begin{equation}
\mathcal{G}(B,\xi):
\Bigg\{
\begin{array}{ccl}
\vett{x} & \rightarrow & \vett{x}' = B (\vett{x} + \vett{\xi}) \\
\grid    & \rightarrow & {\grid'} = \grid B^{-1} \\
\vett{v} & \rightarrow & \vett{v}' = \vett{v} + {\grid} \vett{\xi} 
\end{array}
\Bigg.
\end{equation}
Where $B \in GL \left( N,GF(q) \right)$ (change of basis) and 
$\xi \in {\mathbb{Z}_q}^{N}$ (translation).

This gauge structure is purely a
consequences of the vectorial form of the combination 
$\vett{w}=\grid \vett{x}-\vett{v}$. We can derive a larger gauge group 
from the peculiar form of the Hamiltonian
\[
\ham(\vett{w})= \sum_{i=1}^{M} \left( 1- \delta_q(w_i) \right)
\ef{.}
\]
From this form we deduce that vectors
$\vett{w}_1$, $\vett{w}_2$, differing
only for a permutation of the elements, and for the product of each entry for
a non null value, have the same energy.
The group $\{ 1, 2, \ldots , q-1 \}$, with the product of $GF(q)$ ($q$ a prime
number), is called $\mathbb{Z}_q^*$. We have $M$ copies acting independently on
the $M$ entries of $\vett{w}$. Then we have the permutation group of $M$
elements, $\mathfrak{S}_M$. The group whose elements are a permutation of the
$M$ entries,
followed by the product of each entry for an element of $\mathbb{Z}_q^*$, 
is the group\footnote{
The symbol
$\ltimes$ means that 
$\left( \mathbb{Z}_q^* \right)^{\otimes M}$ is a normal subgroup
of the whole group}
\[
\left( \mathbb{Z}_q^* \right)^{\otimes M} \ltimes \mathfrak{S}_M
\]
We call $\Perm \left( M,GF(q) \right)$ the set of $M \times M$ matrices which
is the $M$-dimensional representation of this group (that is, the matrices
with exactly one non zero element in each row and in each column. Note that
the normal subgroup is the subgroup of diagonal matrices). 
So the most general gauge
transformation $\mathcal{G}$ is
\begin{equation}
\mathcal{G}(S,B,\xi):
\Bigg\{
\begin{array}{ccl}
\vett{x} & \rightarrow & \vett{x}' = B (\vett{x} + \vett{\xi}) \\
\grid    & \rightarrow & {\grid'} = S \grid B^{-1} \\
\vett{v} & \rightarrow & \vett{v}' = S \vett{v} + S{\grid} \vett{\xi} 
\end{array}
\Bigg.
\end{equation}
Where $S \in \Perm \left( M,GF(q) \right)$ (permutation and product on the
entries of $\vett{w}$), 
$B \in GL \left( N,GF(q) \right)$ (change of basis) and 
$\xi \in {\mathbb{Z}_q}^{N}$ (translation).

In the case of the unfrustrated model, in which we impose 
the vector $\vett{v}$
to be the null vector, we will restrict to $\xi=\vett{0}$ gauge
transformations.

A common tool when dealing with systems with quenched disorder is the replica
trick. This method leads to calculate annealed averages over a replicated
system, that is, if the number of replicas is $m$, 
the space of configurations is
the (tensor) product of $m$ copies of the original one, and the Hamiltonian
of the
replicated system $\ham^{(m)}$ is
the sum of the Hamiltonians calculated on the different configurations of each
replica:
\begin{equation}
\ham^{(m)}_{\grid}(\vett{x}_1, \ldots, \vett{x}_m) =
\sum_{a=1}^{m}
\ham_{\grid} (\vett{x}_a)
\ef{.}
\end{equation}
We are interested to the ``annealed'' partition function of this system,
that is the sum of the Gibbs factor 
$e^{- \beta \ham^{(m)}_{\grid}(\vett{x}_1, \ldots, \vett{x}_m) }$
on all the configurations $(\vett{x}_1, \ldots, \vett{x}_m)$ 
{\em and}
all the instances $\grid$.

In this frame, and in our peculiar case, a new symmetry
arises between replicas and clauses, which enlarges the gauge group.
If we call
\[
{x_j}^{(a)} = \widetilde{\grid}_{ja} \ef{;}
\]
and defining
\[
W_{ia}=\sum_{j=1}^{N} \grid_{ij} \widetilde{\grid}_{ja}
\ef{;}
\]
we have for the Hamiltonian the symmetric expression
\begin{equation}
\ham^{(m)}(W)
=
\sum_{a=1}^{m}
\sum_{i=1}^{M}
\left( 1- \delta_q(W_{ia}) \right)
\ef{.}
\end{equation}
From this expression we see that we could permutate the $mM$ elements of the
matrix $W$ in an arbitrary way, and multiply each of them for an independent
element in ${\mathbb{Z}_q}^*$. Such a general class of transformations 
has not a clear action on the physical variables $(\grid, \widetilde{\grid})$.
We could restrict ourselves to the action of a permutation matrix for the
clauses
$S \in \Perm \left( M,GF(q) \right)$ on the left, and a permutation matrix
for the replicas
$S' \in \Perm \left( m,GF(q) \right)$ on the right.
So we have that, as a consequence of
\[
\ham^{(m)}(W) = \ham^{(m)}(SWS')
\]
the general gauge transformation is of the form 
$\mathcal{G}$:
\begin{equation}
\mathcal{G}(S,B,S'):
\Bigg\{
\begin{array}{ccl}
\grid^T  & \rightarrow & {\grid'}^T = S \grid^T B^{-1} \\
\widetilde{\grid} & \rightarrow & \widetilde{\grid}' = 
B \widetilde{\grid} S'
\end{array}
\Bigg.
\end{equation}
Where $S \in \Perm \left( M,GF(q) \right)$ (permutation and product on the
clauses), 
$B \in GL \left( N,GF(q) \right)$ (change of basis on the variables space) 
and 
$S' \in \Perm \left( m,GF(q) \right)$ (permutation and product on the
replicas).

\subsection{The clause-variable (C-V) duality}

For the unfrustrated model we can 
state a notion of duality which formalizes the
heuristic arguments of section \ref{subsec7452} on the 
symmetry of roles between
clauses and variables. We will call this duality the {\em C-V duality}.

The C-V duality maps a $M \times N$ instance of the problem in a $N \times M$
instance.
The matrix $\grid'$ associated to the new instance is simply the transposed of
the original one $\grid'=\grid^T$. As the duality
brings the density parameter $\gamma$ in $\gamma^{-1}$, it has a ``fixed
point'' in this parameter space at $\gamma_c=1$. We will see that this density
is also the critical density for the SAT-UNSAT transition.

In a graphical representation 
the dual  of a generic configuration is a generic
hypergraph, and the dual of a solution is a hyperloop.

More generally, we can consider all the degeneracy numbers of the energy
levels in the two problems
\begin{align}
S^{(n)} &= \sum_{\sigma} \deltakr (\ham_{\grid}(\sigma), n) \ef{;}
&
H^{(m)}&= \sum_{\tau} \deltakr (\ham_{\grid^T}(\tau), m) \ef{;}
\end{align}
which will define a sort of two dual partition functions
associated to each instance
\begin{align}
Z(\grid)&=\sum_n S^{(n)} e^{-\beta n} \ef{;}
&
Z^{(T)}(\grid)&=\sum_m H^{(m)} e^{-\beta m} \ef{.}
\end{align}
A first easy statement deriving from duality could be
\begin{align}
Z(\grid^T) &= Z^{(T)}(\grid) \ef{;}
&
Z^{(T)}(\grid^T) &= Z(\grid) \ef{.}
\end{align}
But the duality provides much stronger relations between
arbitrary functions in the operators $S^{(n)}$ and $H^{(m)}$:
\begin{equation}
f \big( S^{(n)}(\grid) , H^{(m)}(\grid) \big) =
f \big( H^{(n)}(\grid^T) , S^{(m)}(\grid^T) \big) \ef{.}
\end{equation}
And then, averaging over the ensemble of instances:
\begin{align}
\label{7461}
\reval{f \big( S^{(n)},H^{(m)} \big)}_{N, M}&=
\reval{f \big( H^{(n)},S^{(m)} \big)}_{M, N}
&
&\textrm{finite-size;}
\\
\reval{f \big( S^{(n)},H^{(m)} \big)}_{\gamma}&=
\reval{f \big( H^{(n)},S^{(m)} \big)}_{\gamma^{-1}}
&
&\textrm{infinite size limit.}
\end{align}
In the following we will often 
omit the superscript in $S^{(0)}$ and $H^{(0)}$, referring to them simply as
$S$ and $H$.

\subsection{Expectation values of $S$ and $H$ moments}

In what follows we will give elementary
combinatorial derivations of the expectation values for a certain
set of polynomials in $S$ and $H$.

As the 
formal definition of the number of solutions $S$ is
\begin{displaymath}
S(\grid)= \sum_{\sigma} \deltakr (\ham_{\grid}(\sigma),0) \ef{,}
\end{displaymath}
a nice feature of this operator is that averages over $\sigma$ and
over $\grid$ easily commute, so that
\begin{equation}
\label{asd1}
\reval{S}=
\reval{ \sum_{\sigma} 
\deltakr (\ham_{\grid}(\sigma),0) }
=\sum_{\sigma} \reval{
\deltakr (\ham_{\grid}(\sigma),0) } \ef{.}
\end{equation}
We will call $\mathcal{N}_{k}(q^g)$ the number of subsets of order $k$
of linearly independent elements in a group ${\mathbb{Z}_q}^{g}$.
This quantity can be deduced from a simple inductive argument:
if we already have $k-1$ linearly
independent elements, we can choose the $k$-th one 
between the $q^{g}$ elements of the whole ${\mathbb{Z}_q}^{g}$, unless
it is already one of the $q^{k-1}$ elements of
the ${\mathbb{Z}_q}^{k-1}$ generated subgroup, so we must have
\begin{displaymath}
\mathcal{N}_{k}(q^g)=\mathcal{N}_{k-1}(q^g) \cdot (q^g-q^{k-1}) \ef{.}
\end{displaymath}
For the case $k=0$,
we have only the empty set, so $\mathcal{N}_{0}(q^g)=1$. 
This implies:
\begin{equation}
\label{linind}
\mathcal{N}_{k}(q^g)=
\prod_{i<k} (q^g-q^i) \ef{.}
\end{equation}
We will 
introduce a short notation for the recurrent products:
\begin{align}
\label{rab}
\erre{a,b} & =\prod_{j=a+1}^{b} (1-q^{-j}) 
\ef{;}
&
\gausspol{n}{m} &= \frac{\erre{n-m,n}}{\erre{0,m}}=
\frac{\erre{0,n}}{\erre{0,m} \erre{0,n-m}} 
\ef{;}
\end{align}
so that $\mathcal{N}_{k}(q^g)=q^{kg} \erre{g-k,g}$. The expressions
$\gausspol{n}{m}$ are
known in literature as ``Gauss polynomials''; in this specific case they are
polynomials in the variable $q^{-1}$ (see \ref{app1}).

A nice feature of our ensemble 
is that a change of basis doesn't affect the probability
distribution of the ensemble of $\grid$ grids, as it is flat over all the
possible grids.

Given a certain set of $k$ linearly independent configurations 
$\{\sigma^{(1)}, \ldots ,\sigma^{(k)}\}
$, we can calculate 
the probability that all of them are solutions of a given
instance, averaged over the instances.

Using the stability of the ensemble 
under change of basis, we could pass to a new basis in
which the first $k$ elements are the $\sigma^{(1)}, \ldots ,\sigma^{(k)}$. As
they are linearly independent, this change of basis is not singular. It is
easily seen that they are all solutions of the instance if and only
if the first $k$ rows are filled with zeroes, so we have
\begin{equation}
\label{asd2}
\reval{ \prod_{i=1, \cdots , k}
\deltakr (\ham_{\grid}(\sigma^{(i)}),0) }
=
\left( q^{-M}\right)^k \ef{.}
\end{equation}
Collecting equations (\ref{asd1},\ref{linind},\ref{asd2}) we get
\begin{equation}
\label{prodS}
\begin{split}
\reval{\prod_{i<k}(S-q^i)} &=
\reval{\mathcal{N}_k(S)} =
\reval{ \sum_{
\{\sigma^{(1)}, \ldots ,\sigma^{(k)}\}
} \prod_i
\deltakr (\ham_{\grid}(\sigma^{(i)}),0) }
\\
&=
\sum_{
\{\sigma^{(1)}, \ldots ,\sigma^{(k)}\}
}
\reval{ \prod_i
\deltakr (\ham_{\grid}(\sigma^{(i)}),0) }
\\
&=
\mathcal{N}_k(q^N) \,
q^{-kM}
=q^{k(N-M)} \erre{N-k,N} \ef{.}
\end{split}
\end{equation}
The dual of equation (\ref{prodS})
is:
\begin{equation}
\label{prodH}
\reval{\prod_{i<h}(H-q^i)}=
\reval{\mathcal{N}_h(H)}
=\mathcal{N}_h(q^{M}) \,
q^{-h N}
=q^{-h(N-M)} \erre{M-h, M} \ef{.}
\end{equation}
A more complex question we can answer is the probability that $k$ linearly 
independent configurations 
$\{\sigma^{(1)}, \ldots ,\sigma^{(k)}\}
$
are solutions of a certain instance $\grid$, and $h$ linearly 
independent configurations 
$\{\tau^{(1)}, \ldots ,\tau^{(h)}\}
$
are solutions of its dual instance $\grid^T$, at the same time. With a
similar ``double'' change of basis we find that both the first
$k$ rows and the first $h$ columns should be filled with zeroes, so
\begin{equation}
\label{asd3}
\reval{ \prod_{i=1, \cdots , k}
\deltakr (\ham_{\grid}(\sigma^{(i)}),0)
\prod_{j=1, \cdots , h}
\deltakr (\ham_{\grid^T}(\tau^{(j)}),0) }
=
q^{kh -k M -h N} \ef{.}
\end{equation}
If we collect equations (\ref{asd1},\ref{linind},\ref{asd3}) we obtain that
\begin{equation}
\label{prodSH}
\begin{split}
\reval{\prod_{i<k}(S-q^i) \prod_{j<h}(H-q^j)}&=
\reval{\mathcal{N}_k(S) \mathcal{N}_h(H)}
=\mathcal{N}_k(q^{N}) \, \mathcal{N}_h(q^{M})\,
q^{kh -k M -h N}
\\
&=q^{kh} \, q^{(k-h)(N-M)} \, \erre{N-k, N} \erre{M-h, M} \ef{.}
\end{split}
\end{equation}
Notice that $\erre{N-k,N}$ and $\erre{M-h, M}$ are 
exponentially depressed finite-size correction
in the thermodynamic limit. It is also worth noticing that relation
(\ref{prodSH}) for $h=k=1$ reads
\[
\reval{(S-1)(H-1)} = q (1-q^{-N})(1-q^{-M}) \simeq q
\ef{,}
\]
that is, the average of the product between the numbers of 
non-trivial solutions and of
non-trivial hyperloops, in the thermodynamic limit, tends to a constant,
independent from the variable-to-clauses ratio $\gamma=N/M$. This is the
precise formalization of the heuristic solution--hyperloop correspondence of
section \ref{subsec7452}.

\subsection{Expectation values for excited states degeneracy.}

In a way similar to the previous calculation, we could find the
expectation value for expressions involving the number of excited
states $S^{(n)}$, with $n$ violated clauses, or by duality, the
number of
hypergraphs $H^{(n)}$ with $n$
variables appearing in the associated equation.

A quantity which has been considered important in the
literature \cite{rwz} is $H^{(1)}$, as such a kind of hypergraph fixes the
value of a variable, which will be part of the so called ``backbone'',
defined as the subset of variables which take the same value on each
solution. As one trivial solution is the ferromagnetic one, the size
of the backbone is in some sense the magnetization of the system.

The average probability that a certain variable configuration is a
$n$-level excited state 
is exactly $\binom{M}{n} (q-1)^{n} q^{-M}$, 
as, after the change of basis which makes this configuration the first basis
vector, it will be a $n$-level excited state if and only if we have exactly
$n$ non-zero entries in the first column of the grid:
\begin{align}
\reval{ S^{(n)} }
&= \binom{M}{n} (q-1)^n \left( q^{(N-M)} -q^{-M} \right) \ef{;}
\\
\reval{ H^{(n)} }
&= \binom{N}{n} (q-1)^n \left( q^{(M-N)} -q^{-N} \right) \ef{.}
\end{align}
In particular the expression for $\reval{ H^{(1)} }$ 
shows analytically the conjecture,
supported by numerical simulations on traditional
hyper-SAT~\cite{rwz}, that, by varying the density of clauses, 
level 1 hypergraphs arise at the same threshold of hyperloops, 
but in a discontinuous way.

\section{Derivation of $p_{N,M}(R)$ from integer moments of $S$}
\label{probdistrmom}

In this section we will calculate the exact probability distribution 
$p_{N,M}(R)$
for the
rank of $M \times N$ matrices on $GF(q)$, first by calculating the expectation
value of 
integer moments of the number of solutions for the related homogeneous system
$\grid \vett{x} = \vett{0}$, then, solving the ``moments
problem'' (see 
\cite{momenta})
deriving from them.

The order parameter for the SAT/UNSAT phase transition is the probability that
an instance has no other solution beyond the imposed one, that is, the
probability that the rank
$R$ of the matrix $\grid$ is equal to the number of variables $N$, so it is
simply the value of $p_{N,M}(N)$.

\subsection{Calculation of $\reval{S^k}$}

Equation (\ref{prodS}), in the infinite size limit, reads
\begin{equation}
\label{gue}
\reval{\prod_{i<k}(S-q^i)} = q^{-k \Delta} \ef{;}
\end{equation}
where, as we said above, $\Delta = M-N$. 
Given the linear system of the first $k$
equations, we could try to solve it, 
and find an expression for each moment
$\reval{S^k}$. The linear system is:
\begin{equation}
\sum_h A_{kh} \reval{S^h} = q^{-k \Delta} \ef{;}
\end{equation}
where the entries of the matrix $A_{kh}$ are defined by
\[
\prod_{h=0}^{k-1} (S-q^h) = \sum_{h=0}^{k} A_{kh} S^h \ef{.}
\]
This product is analogous to the expression of theorem \ref{thapp1} in
\ref{app1}, so we
obtain
\[
\prod_{h=0}^{k-1} (S-q^h) =(-1)^k q^{\frac{k(k-1)}{2}} 
\prod_{h=0}^{k-1} (1-S q^{-h}) = \sum_{h=0}^{k} 
(-1)^{k-h} S^h q^{\frac{k(k-1)}{2} - \frac{h(h-1)}{2}} 
\gausspol{k}{h}
\]
and, from the definition, we recognize:
\begin{equation}
\label{akh}
A_{kh}=(-1)^{k-h} q^{\frac{k(k-1)}{2} - \frac{h(h-1)}{2}} 
\gausspol{k}{h} \ef{.}
\end{equation}
Note that $A_{kh}$ is a triangular matrix, as obvious, as in the $i$-th
equation only moments up to $\reval{S^i}$ appears. This property is encoded in
equation (\ref{akh}), as in $\erre{k-h,k}$ a zero factor appears when $h>k$.

If we could find the inverse matrix $B_{lk}$ such that
\begin{equation}
\label{8715}
\sum_{k=0}^{n}
B_{lk} A_{kh}=
\delta_{lh}
\qquad n \geq l,h
\ef{,}
\end{equation}
we would solve the system.
We have an ansatz for this matrix:
\begin{ansatz}
\begin{equation}
\label{Bij}
B_{lk}= 
q^{k(l-k)}
\gausspol{l}{k} \ef{.}
\end{equation}
\end{ansatz}
We should verify that
\begin{equation}
\sum_{k=h}^{l}
q^{k(l-k)}
\gausspol{l}{k}
\cdot
(-1)^{k-h}
q^{\frac{k(k-1)}{2} - \frac{h(h-1)}{2}}
\gausspol{k}{h}
= \delta_{lh} \ef{.}
\end{equation}
When $l<h$ this is trivial, as we have no summands. When $l=h$ it is still
easy, as we have only one summand, which is trivially 1. 
When $l>h$ we need a bit of algebra:
\begin{equation*}
\sum_{k=h}^{l}
q^{k(l-k)}
\gausspol{l}{k}
\ (-1)^{k-h}
q^{\frac{k(k-1)}{2} - \frac{h(h-1)}{2}}
\gausspol{k}{h}
=
\sum_{i=0}^{j}
q^{hj}
\gausspol{j+h}{h}
q^{-\frac{i(i-1)}{2}} \left( -q^{j-1} \right)^i
\gausspol{j}{i}
\end{equation*}
where $i=k-h$ and $j=l-h$.
We can recognize again the expression of theorem \ref{thapp1}, 
so the last quantity is
\begin{equation}
\sum_{i=0}^{j}
q^{-\frac{i(i-1)}{2}} \left( -q^{j-1} \right)^i
\gausspol{j}{i}
=\prod_{n=0}^{j-1} (1-q^{j-1} q^{-n}) = 0 \ef{.}
\end{equation}
So, at the end we found the matrix which solves the linear system:
\begin{equation}
\label{fyui}
\reval{S^k}=B_{kh} q^{-h \Delta} 
=\sum_{h=0}^{k}
q^{-h \Delta} 
q^{h(k-h)}
\gausspol{k}{h} \ef{.}
\end{equation}

\subsection{Derivation of $p_{\Delta}(g)$}

If we write the expression for $B_{kh}$ as a polynomial calculated in $q^k$, 
we have
\begin{align}
\label{3822}
\mathcal{B}_h(x)&=\prod_{i=0}^{h-1} \frac{x-q^i}{q^h-q^i} \ef{;}
&
B_{kh}&= \mathcal{B}_h(q^k) \ef{.}
\end{align}
As, by definition, $\reval{S^k}=\reval{(q^g)^k}=\sum_g p(g) (q^k)^g$, we have
\begin{equation}
\label{fyuio}
\sum_g p_{\Delta}(g) (q^k)^g = \sum_{h=0}^{k} q^{-h \Delta} \mathcal{B}_h(q^k)
\ef{;}
\end{equation}
furthermore, as $\mathcal{B}_h(q^k)=0$ for $h >k$, we can extend summation to
infinity. In this form, we could make an ansatz of analytical extension of
the property above, and write
\begin{ansatz}
\label{ansatzinf}
\begin{equation}
\sum_g p_{\Delta}(g) x^g = \sum_{h=0}^{\infty} 
q^{-h \Delta} \mathcal{B}_h(x) \ef{.}
\end{equation}
\end{ansatz}
If this property is formally true, and not only valid for the discrete set of
points $x=q^k$, we could find $p_{\Delta}(g)$ just comparing the series
coefficients. Again from theorem \ref{thapp1} in \ref{app1}, we have
\begin{gather}
\nonumber
\prod_{i=0}^{h-1} (x-q^i) =(-1)^h q^{\frac{h(h-1)}{2}} 
\prod_{i=0}^{h-1} (1-x q^{-i}) = \sum_{g=0}^{h} 
(-1)^{h-g} x^g q^{\frac{h(h-1)}{2} - \frac{g(g-1)}{2}} 
\gausspol{h}{g} \ef{;}
\\
\nonumber
\prod_{i=0}^{h-1} \frac{1}{q^h-q^i}= 
q^{-h^2} \frac{1}{\erre{0,h}} \ef{;}
\end{gather}
so, calling $l=h-g$, we have
\begin{equation}
p_{\Delta}(g) =\frac{q^{g(g+\Delta)}}{\erre{0,g}} 
\sum_{l=0}^{\infty} q^{-\frac{l(l-1)}{2}} (-q^{-(g+\Delta +1)})^l
\frac{1}{\erre{0,l} }
\ef{;}
\end{equation}
with a further use of theorem \ref{thapp1} we finally have
\begin{equation}
\label{9573}
p_{\Delta}(g) =
q^{-g(g+\Delta)} \frac{\erre{0,\infty}}{\erre{0,g} \erre{0,g+\Delta}} \ef{.}
\end{equation}
Note that, as asymptotically the moments diverge faster than exponentially
($\reval{S^k} \sim C \, q^{(k-\Delta)^2/4}$), the
heuristic ansatz \ref{ansatzinf} has no mathematical justification
(see 
\cite{momenta}).

\subsection{Finite-size corrections}

In this procedure it is not difficult to introduce finite-size corrections. In
fact, if we go back to the exact form of equation (\ref{prodS}), we could
write
\begin{equation}
\label{gue2}
\reval{\prod_{i<k}(S-q^i)} = q^{-k \Delta} \erre{N-k,N} \ef{;}
\end{equation}
so the linear system has the same matrix $A_{kh}$, but a different
inhomogeneous term:
\begin{equation}
A_{kh} \reval{S^h} = q^{-k \Delta} \erre{N-k,N} \ef{.}
\end{equation}
All the procedure of inversion of $A_{kh}$ are still valid, and we just need
to replace equations (\ref{fyui}, \ref{fyuio}) with the correct
\begin{equation}
\reval{S^k}=
\sum_g p_{\Delta}(g) (q^k)^g = \sum_{h=0}^{k} q^{-h \Delta}
\erre{N-h,N} \mathcal{B}_h(q^k) \ef{.}
\end{equation}
Now, beyond a finite-size cutoff $k \gg 2N+\Delta$ (as 
$\erre{N-h,N}=0$ for $h>N$), moments diverge like
$\reval{S^k} \sim C \, q^{N (k-N-\Delta)}$, and the finite-size version of the
ansatz \ref{ansatzinf} is actually a precise mathematical statement:

\begin{ansatz}
\begin{equation}
\sum_g p_{\Delta}(g) x^g = \sum_{h=0}^{\infty} 
q^{-h \Delta} \erre{N-h,N} \mathcal{B}_h(x) \ef{.}
\end{equation}
\end{ansatz}
This is a rare case in which
a careful
finite-size regularization 
demonstrates the validity of a heuristic analytic continuation.

Now just some algebra follows: we can correct equation (\ref{9573}) in
\begin{equation}
p_{N,M}(g) =\frac{q^{g(g+\Delta)}}{\erre{0,g}} 
\sum_{l=0}^{N-g} q^{-\frac{l(l-1)}{2}} (-q^{-(g+\Delta +1)})^l
\frac{\erre{N-g-l,N}}{\erre{0,l} }=
q^{-g(g+\Delta)} \frac{\erre{0,N+\Delta} \erre{N-g,N}}
{\erre{0,g} \erre{0,g+\Delta}}
\end{equation}
Introducing the rank of the matrix $R=N-g$ to 
symmetrize the notations:
\begin{equation}
\label{cghjR}
p_{N,M}(\rrank (\grid) = R) = q^{-(N-R)(M-R)} 
\erre{0,R}
\gausspol{N}{R}
\gausspol{M}{R} \ef{.}
\end{equation}

\subsection{Finite-size scaling function of the SAT/UNSAT order parameter}
\label{subsec897}

Equation (\ref{cghjR}) provides us all the informations about the behaviour of
our model, at zero temperature with respect to the fictitious Hamiltonian 
(\ref{hamiltx}). If we want to compare our results with the standard ``open
questions'' of random satisfiability problems, we should concentrate on the
order parameter for the SAT/UNSAT transition, that is
\[
\phi (\gamma, M) =
p_{N,M}(\rrank(\grid)=N)
\ef{,}
\]
where it is understood $N=\gamma M$.
The results of equation (\ref{cghjR}) leads to
\[
\phi (\gamma, M) =
\erre{0,N} \gausspol{M}{N}
\ef{.}
\]
\begin{figure}[!tb]
\begin{center}
\label{fig451}
\includegraphics[width=279pt]{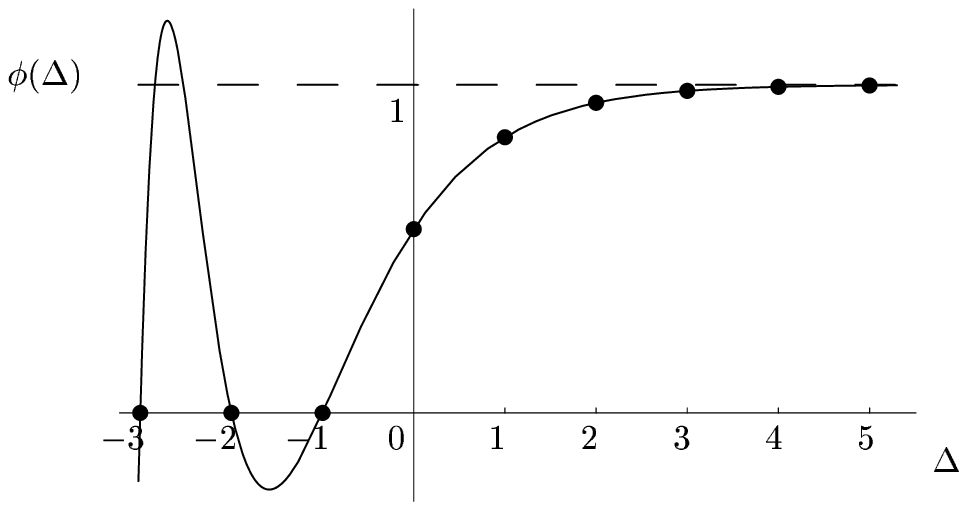}
\caption{Order parameter $\phi(\Delta)$ as a function of 
  $\Delta =M-N$, for
  $q=3$. The ``physical points'' are for integer values of $\Delta$, the
  continuos curve corresponds to the analytical continuation (\ref{53286}).}
\end{center}
\end{figure}
The
thermodynamic limit $M \rightarrow \infty$, 
at fixed density $\gamma$, allows to determine the
critical density $\gamma_c=1$:
\begin{equation}
\phi (\gamma, \infty) =
\lim_{M \rightarrow \infty}
\frac{\erre{0,M}}{\erre{0,(1-\gamma)M}}
=
\left\{
\begin{array}{ll}
1               & \gamma < 1 \\
\erre{0,\infty} & \gamma = 1 \\
0               & \gamma > 1
\end{array}
\right.
\ef{.}
\end{equation}
We expect a non trivial $M \rightarrow \infty$ limit when we keep fixed a
certain combination of density and size of the system, 
$(\gamma-\gamma_c) M^{1/\nu}$.
Again from equation (\ref{cghjR}) it is clear that the proper scaling window
is obtained for $\nu=1$.
Introducing 
the parameter $\Delta = (\gamma_c - \gamma) M$,
and performing
the thermodynamic limit at fixed $\Delta$, we obtain back 
the result implicit in
equation (\ref{9573}):
\begin{equation}
\label{90821}
\phi(\Delta)=
\lim_{M \rightarrow \infty}
\frac{\erre{0,M}}{\erre{0,\Delta}}
= 
\frac{\erre{0,\infty}}{\erre{0,\Delta}}
\ef{.}
\end{equation}
Although
this result makes sense 
only for $\Delta$ an integer number,
analytical
extension to $\Delta$ an arbitrary complex number 
is possible if we use the property
\[
\erre{0,n}=\frac{\erre{0,\infty}}{\erre{n,\infty}}
\ef{,}
\]
to restate equation (\ref{90821}) as
\begin{equation}
\label{53286}
\phi(\Delta)
= \erre{\Delta, \infty} = \prod_{k=1}^{+ \infty}
(1-q^{-\Delta - k})
\ef{.}
\end{equation}
A plot of this function for a typical value of $q$, and $\Delta$ a real
 number, is shown in figure.

\section{Derivation of $p_{N,M}(R)$ from Gauss algorithm}
\label{probdistr}

As we have seen in section \ref{subsec6598},
we can perform a gauge transformation which, on the space of configurations,
is essentially a change of basis. 
We can perform a gauge fixing to make clear the rank of a instance matrix
$\grid$, and, from the analysis of this procedure, rederive the probability
distribution for the rank, $p_{N,M}(R)$, which we already found in a
purely algebraic way in equation (\ref{cghjR}).

We will define our gauge fix in an algorithmic way, which we call 
{\em Gauss algorithm}, as it is essentially the standard algorithm to
triangulate a matrix. We remark that
we should only admit ``moves'' on the columns, as they correspond to a 
change of basis in the space of configurations 
($\grid \rightarrow \grid B^{-1}$, with
$B \in GL(N)$),
while we only admit simple permutations
on the rows
($\grid \rightarrow S \grid$, with
$S \in \Perm(M)$).

\subsection{Step zero}

Gauss algorithm takes a $M \times N$ matrix $\grid$, and reduce it
in a ``normal form''
\begin{equation}
\grid_{\textrm{norm}}=
\begin{array}{|ccccccc|}
\hline
1 & \multicolumn{3}{c|}{} & 0 & 0 & \cdot \\ \cline{1-1}
  & \multicolumn{1}{|c}{1} & 
\multicolumn{2}{c|}{0} & 0 &&
\\
\cline{2-2}
\multicolumn{2}{|c|}{}&\cdot &
\multicolumn{1}{c|}{} & 0 && \\
\cline{3-3}
\multicolumn{3}{|c|}
{}
&\multicolumn{1}{|c|}{1} & \cdot && \\
\cline{4-4}
\multicolumn{3}{|c}{\grid_{\textrm{red}}} & & \multicolumn{3}{|c|}{} \\
\multicolumn{4}{|c|}{} & \multicolumn{3}{|c|}{} \\
\multicolumn{4}{|c|}{} & \multicolumn{3}{|c|}{} \\
\hline
\end{array}
\end{equation}
where the central triangular matrix has side $R=N-g$,
corresponding to the rank of the matrix.

Suppose we know the probability distribution $p_{N,M}(R)$. Now consider
the two following inductive steps:

\subsection{Induction on columns (add one variable)}
\label{indcol}

Take a $M \times (N+1)$ matrix $\grid$, and apply the Gauss algorithm
neglecting the last column $c$. The new matrix will be in a ``pre-normal form''
\begin{equation}
\label{pnf1200}
\grid_{\textrm{norm}}^{(1)}=
\begin{array}{|ccccccc|c|}
\hline
1 & \multicolumn{3}{c|}{} & 0 & 0 & \cdot & \\ \cline{1-1}
  & \multicolumn{1}{|c}{1} & 
\multicolumn{2}{c|}{0} & 0 && &
\\
\cline{2-2}
\multicolumn{2}{|c|}{}&\cdot &
\multicolumn{1}{c|}{} & 0 &&& \\
\cline{3-3}
\multicolumn{3}{|c|}
{}
&\multicolumn{1}{|c|}{1} & \cdot &&& c \\
\cline{4-4}
\multicolumn{3}{|c}{\grid_{\textrm{red}}} & & \multicolumn{3}{|c|}{} & \\
\multicolumn{4}{|c|}{} & \multicolumn{3}{|c|}{} & \\
\multicolumn{4}{|c|}{} & \multicolumn{3}{|c|}{} & \\
\hline
\end{array}
\end{equation}
(note that the probability distribution for the last column is flat). 
Now reduce
to zero the first $R$ terms of the column via Gauss algorithm, obtaining
\begin{equation}
\label{pnf1201}
\grid_{\textrm{norm}}^{(2)}=
\begin{array}{|ccccccc|c|}
\hline
1 & \multicolumn{3}{c|}{} & 0 & 0 & \cdot & 0 \\ \cline{1-1}
  & \multicolumn{1}{|c}{1} & 
\multicolumn{2}{c|}{0} & 0 && & 0
\\
\cline{2-2}
\multicolumn{2}{|c|}{}&\cdot &
\multicolumn{1}{c|}{} & 0 &&& \cdot \\
\cline{3-3}
\multicolumn{3}{|c|}
{}
&\multicolumn{1}{|c|}{1} & \cdot &&& \\
\cline{4-4} \cline{8-8}
\multicolumn{3}{|c}{\grid_{\textrm{red}}} & & \multicolumn{3}{|c|}{} & \\
\multicolumn{4}{|c|}{} & \multicolumn{3}{|c|}{} & c' \\
\multicolumn{4}{|c|}{} & \multicolumn{3}{|c|}{} & \\
\hline
\end{array}
\end{equation}
(note that the probability distribution for the remaining part of the 
column is
still flat!). Now we have two ways to get further:
\begin{description}
\item[$c'$ contains only zeroes] (this happens with
  probability $q^{-M+R}$): the matrix is already in normal form, as we 
  have simply added an
  empty row, so $R'=R$;
\item[$c'$ contains some non-zero element] (this happens with
  probability $1-q^{-M+R}$): we will bring the column on position $R+1$, and
  eventually permute two of the last $M-R$ rows in order to have a 1 in the
  diagonal term, and again a matrix in normal form. 
  At the end, we have $R'=R+1$;
\end{description}
The recursion law deriving from these reasonings is resumed in the diagram
below:
\[
\xymatrix{
\ldots
\ar[dr]
&
p_{N,M}(R-1)
\ar[d]|{q^{-M+R-1}} \ar[dr]|{1-q^{-M+R-1}}
&
p_{N,M}(R)
\ar[d]|{q^{-M+R}} \ar[dr]|{1-q^{-M+R}}
&
\ldots
\ar[d]
\\
&
p_{N+1,M}(R-1)
&
p_{N+1,M}(R)
&
p_{N+1,M}(R+1)
}
\]
The analysis of the diagram leads to the equation
\begin{equation}
\label{prob1}
p_{N+1,M}(R)=(1-q^{-M+R-1}) p_{N,M}(R-1) 
+ q^{-M+R} p_{N,M}(R) \ef{.}
\end{equation}

\subsection{Induction on rows (add one clause)}
\label{column}

Take a $(M+1) \times N$ matrix $\grid$, and apply the Gauss algorithm
neglecting the last row $r$. The new grid will be in a ``pre-normal form''
\begin{equation}
\grid_{\textrm{norm}}^{(1)}=
\begin{array}{|ccccccc|}
\hline
1 & \multicolumn{3}{c|}{} & 0 & 0 & \cdot \\ \cline{1-1}
  & \multicolumn{1}{|c}{1} & 
\multicolumn{2}{c|}{0} & 0 &&
\\
\cline{2-2}
\multicolumn{2}{|c|}{}&\cdot &
\multicolumn{1}{c|}{} & 0 && \\
\cline{3-3}
\multicolumn{3}{|c|}
{}
&\multicolumn{1}{|c|}{1} & \cdot && \\
\cline{4-4}
\multicolumn{3}{|c}{\grid_{\textrm{red}}} & & \multicolumn{3}{|c|}{} \\
\multicolumn{4}{|c|}{} & \multicolumn{3}{|c|}{} \\
\multicolumn{4}{|c|}{} & \multicolumn{3}{|c|}{} \\
\hline
\multicolumn{7}{|c|}{r} \\
\hline
\end{array}
\end{equation}
We call $r'$ the last $N-R$ terms of the row $r$, in analogy to
the previous section. Note that the probability distribution for these terms is
again flat. Now we have two ways to get further:
\begin{description}
\item[$r'$ contains only zeroes] (this happens with
  probability $q^{-N+R}$): the matrix is already in normal form, with
$R'=R$;
\item[$r'$ contains some non-zero element] (this happens with
  probability $1-q^{-N+R}$): we will bring the row on position $R+1$,
and apply Gauss Algorithm to the last 
$N-R$ rows, in order to have just one 1 left in the first term (the diagonal
one), 
and again a matrix in normal form.
At the end, we have
$R'=R+1$;
\end{description}
Again the recursion law deriving from 
these reasonings is resumed in a diagram:
\[
\xymatrix{
\ldots
\ar[dr]
&
p_{N,M}(R-1)
\ar[d]|{q^{-N+R-1}} \ar[dr]|{1-q^{-N+R-1}}
&
p_{N,M}(R)
\ar[d]|{q^{-N+R}} \ar[dr]|{1-q^{-N+R}}
&
\ldots
\ar[d]
\\
&
p_{N,M+1}(R-1)
&
p_{N,M+1}(R)
&
p_{N,M+1}(R+1)
}
\]
The analysis of this diagram leads to a second equation
\begin{equation}
\label{prob2}
p_{N,M+1}(R)=(1-q^{-N+R-1}) p_{N,M}(R-1) 
+ q^{-N+R} p_{N,M}(R) \ef{.}
\end{equation}

\subsection{Derivation of the probability distribution}

The two equations (\ref{prob1},\ref{prob2}) are linear and homogeneous. The
quantities involved depends on three independent parameters: $N$, $M$
and $R$. We will show ``a posteriori'' that the quantity $p_{N,M}(R)$
has a good thermodynamic limit $M \rightarrow \infty$ when the limit
is performed at $\Delta=M-N$ and $g=N-R$ fixed. We will call this
limit function $p_{\Delta}(g)$.

This assumption allows us to write equations (\ref{prob1},\ref{prob2})
as a system of two linear equations for $p_{\Delta}(g)$, in the two
variables $g$ and $\Delta$:
\begin{equation}
\left\{
\begin{array}{l}
p_{\Delta-1}(g)=
(1-q^{-g-1}) p_{\Delta}(g+1) + q^{-g} p_{\Delta}(g)
\\
p_{\Delta+1}(g)=
(1-q^{-g-1+\Delta}) p_{\Delta}(g+1) + q^{-g+\Delta} p_{\Delta}(g)
\end{array}
\right.
\ef{;}
\end{equation}
An easy to verify solution is
\begin{equation}
p_{\Delta}(g)=p_{0}(0) \frac{q^{-g(g+\Delta)}}
{\erre{0,g} \erre{0,g+\Delta}} \ef{.}
\end{equation}
We can find the right normalization from the limit
\begin{equation}
\lim_{\Delta \rightarrow +\infty} p_{\Delta}(0)=1 \ef{,}
\end{equation}
which leads to
\begin{equation}
\label{probgeq}
p_{\Delta}(g)
=q^{-g(g+\Delta)} \frac{\erre{0,+\infty}}{\erre{0,g} \erre{0,g+\Delta}}
\ef{,}
\end{equation}
which is again equation (\ref{9573}).

This formula was previously (independently) discovered
by the Russian school (see \cite{kova1, kova2}), with a
procedure analogous to the one of this subsection.

\subsection{Finite-size corrections}
\label{subsecFSCN}

In this paragraph we will show how, with a careful use of the theory of
partitions (see \cite{rade}, and, for a summary of the main result,
\ref{app1}), the only equation (\ref{prob1}) (or, in a similar way, the
only equation (\ref{prob2})) leads to the exact expression of
$p_{N,M}(R)$, calculated directly from the initial conditions,
with no need of any assumption.

Equation (\ref{prob1}) was
\begin{equation*}
p_{N+1,M}(R)=(1-q^{-M+R-1}) p_{N,M}(R-1) 
+ q^{-M+R} p_{N,M}(R) \ef{.}
\end{equation*}
It is more suitable to switch to the variable $S=N-R$ instead of $N$ for
some steps:
\begin{equation}
\label{78231}
p_{M}(S,R)=(1-q^{-M+R-1}) p_{M}(S,R-1) 
+ q^{-M+R} p_{M}(S-1,R) \ef{.}
\end{equation}
Imagine to expand recursively this equation until we reconduce $p_M(S,R)$
to the initial conditions
\[
p_M(0,0)=1
\ef{;}
\qquad p_M(S,R)=0 
\quad \textrm{ if } \quad
S<0, R<0
\ef{.}
\]
This would produce a prolification of terms. Each of these terms can be
labeled by a trajectory on a $S \times R$ grid, starting from the
point $(0,0)$ on the top left corner, and stopping at the point
$(S,R)$ on the bottom right corner. This trajectory is directed:
we only admit steps on the right nearest neighbour 
$(s,r) \rightarrow (s,r+1)$ or to the down nearest neighbour
$(s,r) \rightarrow (s+1,r)$. The weights of these steps, determined by
equation (\ref{78231}), are
\begin{equation*}
\xymatrix{
(s,r) 
\ar[r]^{1-q^{-M+r}}
\ar[d]^{q^{-M+r}}
&  (s,r+1) \\
(s+1,r) & }
\end{equation*}
An interesting fact about these weights is that they only depend on
$r$. For this reason, the contribute deriving from the horizontal
steps is the same in all the trajectories, being
\begin{equation}
\label{3658}
\prod_{r=0}^{R-1} (1-q^{-M+r}) = \erre{M-R,M}
= \erre{0,R} \gausspol{M}{R}
\ef{.}
\end{equation}
The contribution deriving from the vertical steps is more complicated,
and needs some results from the theory of partitions. Consider the
trajectory as a string of $S$ integer numbers $(r_1, \ldots, r_S)$,
corresponding to the length of the rows of the diagram individuated by
the trajectory. As the trajectory is directed, we should have 
$0 \leq r_i \leq r_{i+1} \leq R$.
\[
\begin{array}{r|cccccccc|l}
\cline{2-9}
(0,0)^{\nearrow}\!\!\! & \multicolumn{1}{|c|}{} &&&&&&&& \\
\cline{3-3}
& \multicolumn{2}{|c|}{} &&&&&&&  \\
& \multicolumn{2}{|c|}{} &&&&&&& \\
\cline{4-4}
& \multicolumn{3}{|c|}{\!\!\!\! \leftarrow \! r_i \! \rightarrow \!\!\!\!} &&&&&&  \\
\cline{5-7}
& \multicolumn{6}{|c|}{} &&&  \\
\cline{8-9}
& \multicolumn{8}{|c|}{} & \!\! _{\swarrow} (S,R) \\
\cline{2-9}
\end{array}
\]
The vertical steps contribution for
a certain trajectory will be
$
\prod_{i=1}^{S} q^{-M+r_i}
$.
As a result of theorem \ref{thapp2} in \ref{app1} we could write the
generating function in the number of rows:
\begin{equation}
f(z)=\sum_S \ z^S \!\!\! \sum_{(r_1, \ldots, r_S)} \prod_{i=1}^{S} q^{-M+r_i}
= \prod_{r=0}^{R} \frac{1}{1-z q^{-M+r}}
\ef{,}
\end{equation}
and from theorem \ref{thapp3} we have
\begin{equation*}
\prod_{r=0}^{R} \frac{1}{1-z q^{-M+r}}
= \sum_{k=0}^{\infty} z^k q^{-k(M-R)} \gausspol{R+k}{R}
\ef{,}
\end{equation*}
and, as we are interested in the term $z^S$, for $S=N-R$, we finally have
\[
q^{-(N-R)(M-R)} \gausspol{N}{R}
\]
which, combined with the contribution deriving from the horizontal
steps, formula (\ref{3658}), gives again the result of equation
(\ref{cghjR}).

\section{Conclusions}

We introduced a variant of the hyper-SAT model, by relaxing the
constraint on the length of the clauses. This makes the model exactly
solvable, so that we could not only compute the probability for an
instance of being satisfiable, but also the probability distribution
for the number of configurations which satisfy all the constraints.
This probability distribution can be evaluated also for finite number
of variables $N$, and of clauses $M$.
The SAT/UNSAT transition occurs, in the thermodynamic limit, in which
$N$ and $M$ go to infinity, with their ratio fixed to one. This model
enjoys a dual formulation, in which the role of clauses and variables
is interchanged, in agreement with the previous remark that the phase
transition occurs when the number of clauses is equal to the number of
instances.
The whole analysis is soon extended to the general $GF(q)$ case.
Finite-size scaling functions are easily derived from the general
solutions.



The reasons for our work were mainly three:
\begin{enumerate}
\item Random matrices on $GF(q)$, with $N>M$, are connected to random linear
  codes of the Error Correcting Codes theory, in particular to the Random
  Codeword Model introduced by A.~Montanari in \cite{montan01}. The unexpected
  existence
  for an exact solution of the zero-temperature thermodynamic, and some other
  exact results also for the finite-temperature
  opens new perspectives in the study of this model.
\item The random satisfiability of sparse linear systems 
  on $GF(q)$ is a polynomial
  problem, of computation complexity $N^3$. A careful average-case complexity
  analysis shows both a SAT/UNSAT transition (discontinuity in the order
  parameter ``probability for an instance to have at least one solution'', as
  a function of the number of variables to number of clauses ratio), and 
  a ``dynamical'' transition in the complexity class, from linear to cubic
  complexity (see \cite{blrz}). These characteristics make the hyper-SAT model
  an ideal toy model to investigate the underlying mechanisms of 
  average-case complexity in the $K$-SAT
  problem. Our model is a further idealization of hyper-SAT, where the
  dynamical transition is lost, which has the advantage of an exact
  analytical control, also of the replica approach.
\item The random satisfiability of sparse linear systems 
  on $GF(2)$ can be restated as a combinatorial problem in Random Hypergraphs
  theory. Our results are thus relevant also in this area.
\end{enumerate}
The analysis of our model at finite temperature will be presented elsewhere
\cite{inprep1}.


                       \appendix

\section{Results from theory of partitions.}
\label{app1}

In this appendix we summarize some results from analytic number theory of
partitions, mainly due to the ancient work of Euler and Gauss on the generating
function formalism.
Our main reference for these results is Chapter 12 of H.~Rademaker's textbook
\cite{rade}.

It is worth noticing that the mathematical results of this appendix are the
groundwork for important results of number theory known as
{\em Rogers-Schur-Ramanujan identities}, which have been shown to be connected
to the Virasoro characters of conformal minimal models (see e.g.~the
articles of McCoy {\em et.~al.}, \cite{mccoy1,mccoy2}).
They are also known as {\em $q$-calculus}, and are connected with
one-dimensional asymmetric exclusion processes (ASEP) (see e.g.~the work of
T.~Sasamoto, \cite{asep}).

Given an integer $n$, a \emph{partition} $[p]_n$ is a list of ordered integers
$(n_1, n_2, \ldots , n_k)$
such that their sum is $n$:
\[
[p]_n=(n_1, n_2, \ldots , n_k); \qquad 
1 \leq n_i \leq n_{i+1}; \qquad
\sum n_i =n;
\]
A common tool consists in
introducing formal power series in which the $n$-th series coefficient 
is the sum, on all the partitions $[p]_n$ of the
integer $n$, of a certain weight $W([p]_n)$. In the more general framework, 
the weight function could be
valued in $\mathbb{C}$. In the case of ``well behaved'' 
positive real weights, we can think to the power series as a 
grandcanonical partition function on the set of partitions.

An analytic expression for these power series could be found for a wide class
of weight functions. We will restrict ourselves to weights which factorizes on
the elements of the list, that is, given a certain function
$w: \mathbb{N} \rightarrow \mathbb{C}$,
the weight of a partition is:
\[
W\left( [p]_n \right) = \prod_{i=1}^{k} w(n_i) \ef{.}
\]
We can state
the first easy result as follows:
\begin{theor}
\label{thapp2}
The generating function for the partitions, with a factorized weight 
$W\left( [p]_n \right) = \prod_{i=1}^{k} w(n_i)$, is
\begin{equation}
\sum_n z^n \sum_{[p]_n} W([p]_n)=
\prod_{i=1}^{\infty} \frac{1}{1-z w(i)} \ef{.}
\end{equation}
\end{theor}
In the most common case, in which we just want to count all the partitions of
a certain integer, the proper weight would be $w(n)=q^{-n}$, so we will be
concerned with products of the kind $\prod_i (1-z q^{-i})$. 
It will be useful, 
as in equations (\ref{rab}), to introduce the notation
\begin{equation}
\erre{m,n}=\prod_{j=m+1}^{n} (1-q^{-j}) \ef{;} 
\qquad
\gausspol{n}{m}=\frac{\erre{n-m,n}}{\erre{0,m}} \ef{.}
\end{equation}
A more common notation, but more tedious to handle in our context, is the
following:
\begin{align}
(a;x)_n &= \prod_{i=0}^{n-1} (1-a x^i) \ef{;}
&
\gausspol{n}{m}_x &= \frac{(x;x)_n}{(x;x)_{m} (x;x)_{n-m}} 
\ef{,}
\end{align}
where the quantity on the right is known as the $(n,m)$ Gaussian polynomial
in the variable $x$. In this paper we always used $q^{-1}$ as argument of
$\erre{m,n}$ products. 
A few easy relations between the Gaussian polynomials are:
\begin{subequations}
\begin{gather}
\label{54890a}
\gausspol{n}{m} 
= \gausspol{n}{n-m}
\ef{;}
\\
\label{54890b}
\gausspol{n}{m} =
\gausspol{n-1}{m-1} + q^{-m} \gausspol{n-1}{m}
=
q^{-(n-m)} \gausspol{n-1}{m-1} + \gausspol{n-1}{m}
\ef{;}
\\
\label{54890c}
\gausspol{n}{m} \gausspol{m}{l}
=
\gausspol{n}{l} \gausspol{n-l}{m-l}
\ef{.}
\end{gather}
\end{subequations}
Furthermore, at fixed $m$, in the limit $n \rightarrow \infty$ we
have
\[
\lim_{n\rightarrow \infty} 
\gausspol{n}{m}=\frac{1}{\erre{0,m}} \ef{.}
\]
Now consider the formal power series
\[
P_n(x,z)=\prod_{i=1}^{n} (1-z x^i) = \sum_{m=0}^{n} a_{m,n}(x) z^m \ef{;}
\]
from the relation
\[
P_n(x,x z)= \frac{1-z x^{n+1}}{1-zx} P_n(x,z)
\]
we find that
\[
a_{m,n}(x)= -x^m \frac{1-x^{n-m+1}}{1-x^m} a_{m-1,n}(x)
\]
and from the initial condition $a_{0,n}(x)=1$ we find
\[
a_{m,n}(x)= \prod_{m'=1}^{m} -x^{m'} \frac{1-x^{n-m'+1}}{1-x^{m'}} \ef{,}
\]
from which we recognize
\begin{theor}
\label{thapp1}
\begin{subequations}
\begin{gather}
\prod_{i=1}^{n} (1-z q^{-i}) = \sum_{m=0}^{n} (-z)^m q^{-\frac{m(m-1)}{2}} 
\gausspol{n}{m} \ef{;}
\\
\prod_{i=1}^{\infty} (1-z q^{-i}) = 
\sum_{m=0}^{\infty} (-z)^m q^{-\frac{m(m-1)}{2}} 
\frac{1}{\erre{0,m}} \ef{.}
\end{gather}
\end{subequations}
\end{theor}
Analogously, if we consider the formal power series
\[
Q_n(x,z)=\prod_{i=0}^{n} \frac{1}{1-z x^i} = \sum_{m=0}^{\infty} 
b_{m,n}(x) z^m \ef{;}
\]
from the relation
\[
Q_n(x,x z)= \frac{1-z}{1-z x^{n+1}} Q_n(x,z)
\]
we find that
\[
b_{m,n}(x)= \frac{1-x^{m+n}}{1-x^m} b_{m-1,n}(x)
\]
and from the initial condition $b_{0,n}(x)=1$ we find
\[
b_{m,n}(x)= \prod_{m'=1}^{m} \frac{1-x^{m'+n}}{1-x^{m'}}= \gausspol{m+n}{m}
\ef{,}
\]
from which we recognize
\begin{theor}
\label{thapp3}
\begin{subequations}
\begin{gather}
\prod_{i=0}^{n} \frac{1}{1-z q^{-i}} = \sum_{m=0}^{\infty} 
z^m \gausspol{m+n}{m} \ef{;}
\\
\prod_{i=0}^{\infty} \frac{1}{1-z q^{-i}}
=\sum_{m=0}^{\infty} z^m 
\frac{1}{\erre{0,m}} \ef{.}
\end{gather}
\end{subequations}
\end{theor}

                  \ack 

We are grateful to F.~Ricci-Tersenghi
for many fruitful discussions, and for suggesting reference
 \cite{kolchin}. One of us (A.S.) is also grateful to all the 
``guys of Rome group'' 
for the frequent and kind hospitality.

We are also grateful to A.~Montanari for useful discussions and bibliographic
references on connection between our model and error-correcting codes.

We are grateful to F.~Caravenna for suggesting reference \cite{asep} on
connection of $q$-calculus with ASEP models.

              \references  

\end{document}